%% file: main.tex
\documentclass{bioinfo}
\copyrightyear{2023} \pubyear{2023}
 \pdfoutput=1

\usepackage[numbers]{natbib}

\usepackage{breakcites}
\usepackage{multirow}
\usepackage{url}

\usepackage{setspace}
\usepackage[hang]{footmisc}

\copyrightyear{2023} \pubyear{2023}
\newcommand{\codename}{GateSeeder}

\definecolor{dgreen}{rgb}{0.00, 0.5, 0.00}
\newcommand{\sr}[1]{{\color{black}{#1}}}
\newcommand{\ju}[1]{{\color{black}{#1}}}
\newcommand{\gs}[1]{{\color{black}{#1}}}

\newcommand{\srr}[1]{[{\color{red}\bf{ALSER:} \it{#1}}]}

\usepackage{tikz}
\newcommand*\circled[1]{\tikz[baseline=(char.base)]{
            \node[shape=circle,fill,inner sep=1pt] (char) {\textcolor{white}{#1}};}}

\usepackage{algorithm}
\usepackage{algpseudocode}
\usepackage{lscape}
\usepackage{comment}
\usepackage{graphicx}

\usepackage{newfloat}
\DeclareFloatingEnvironment[fileext=lop]{photo}

\usepackage[colorlinks = true,
            linkcolor = blue,
            urlcolor  = blue,
            citecolor = blue,
            anchorcolor = blue,
            breaklinks
            ]{hyperref}
\usepackage{xcolor}
\definecolor{darkblue}{rgb}{0,0,0.8}

\newcommand{\secref}[1]{\hyperref[sec:#1]{\S\ref*{sec:#1}}}

\renewcommand{\algref}[1]{{\hyperref[alg:#1]{Alg.~\ref*{alg:#1}}}}

\newcommand{\alglineref}[1]{{\hyperref[algline:#1]{Line~\ref*{algline:#1}}}}

\usepackage{xspace}
\usepackage{caption}
\usepackage{float}
\begin{document}


\setstretch{1.3}
\firstpage{1}

\subtitle{}

\title[\codename]{\center{\codename: Near-memory CPU-FPGA Acceleration \\of Short and Long Read Mapping}}
\author[Eudine and Alser \textit{et~al}.]{\center{Julien Eudine\,$^{\text{\sfb 1,}*,\#}$, Mohammed Alser\,$^{\text{\sfb 1,}*,\#}$, Gagandeep Singh\,$^{\text{\sfb 1}}$,\\
Can Alkan\,$^{\text{\sfb 2}}$,
Onur Mutlu\,$^{\text{\sfb 1,}*}$
}}
\address{\center{$^{\text{\sf 1}}$Department of Information Technology and Electrical Engineering, ETH Zurich, Zurich 8006, Switzerland \\
$^{\text{\sf 2}}$Department of Computer Engineering, Bilkent University, Ankara, Turkey.}}

\corresp{$^\ast$To whom correspondence should be addressed.
\newline$^\#$Joint first author.}

\history{}

\editor{}

\abstract{\input{paper/abstract}}

\maketitle

\input{paper/introduction}
\input{paper/methods}

\input{paper/results}

\input{paper/discussion}

\input{paper/funding}

\bibliographystyle{unsrtnat}
\bibliography{main}


\clearpage

\appendix

\input{paper/appendix}

\end{document}

%% file: paper/abstract.tex

\textbf{Motivation:}  Read mapping is a computationally expensive process and a major bottleneck in genomics analyses. The performance of read mapping is mainly limited by the performance of three key computational steps: Index Querying, See\gs{d} Chaining, and Sequence Alignment.
The first step is dominated by how fast and frequent it accesses the main memory (i.e., memory-bound), while the latter two steps are dominated by how fast the CPU can compute their computationally-costly dynamic programming algorithms (i.e., compute-bound). 
Accelerating these three steps by exploiting new algorithms and new hardware devices is essential to accelerate most genome analysis pipelines that widely use read mapping.
Given the large body of work on accelerating Sequence Alignment, 
this work focuses on significantly improving the remaining steps.\\
\textbf{Results:} We introduce \emph{\codename{}}, the \emph{first} CPU-FPGA-based near-memory acceleration of both short and long read mapping.
\codename{} exploits near-memory computation capability provided by modern FPGAs that couple a reconfigurable compute fabric with high-bandwidth memory (HBM) to overcome the memory-bound and compute-bound bottlenecks.
\codename{} also introduces a new lightweight algorithm for finding the potential {matching segment pairs}. 
Using real ONT, HiFi, and Illumina sequences, we experimentally demonstrate that \codename{} outperforms Minimap2, without performing sequence alignment, by up to 40.3$\times$, 4.8$\times$, and 2.3$\times$, respectively.
\sr{When performing read mapping with sequence alignment, \codename{} outperforms Minimap2 by 1.15-4.33$\times$ (using KSW2) and by 1.97-13.63$\times$ (using WFA-GPU).}
\\
\textbf{Availability:} \href{https://github.com/CMU-SAFARI/GateSeeder}{https://github.com/CMU-SAFARI/GateSeeder}\\
\textbf{Contact:} \href{julien@eudine.fr}{julien@eudine.fr}, \href{mealser@gmail.com}{mealser@gmail.com}, \href{omutlu@ethz.ch}{omutlu@ethz.ch}
\\
\textbf{Supplementary information:} Supplementary data are available at \textit{Bioinformatics}
online.

%% file: paper/introduction.tex

\section{Introduction}\label{sec:introduction}

Read mapping is the first fundamental step in most genomic analyses~[\citenum{9154510, ALSER20224579,vaser2017fast, liao2023draft, Alser2021,lapierre2020metalign, meyer2022critical,lapierre2019micop}].
Read mapping compares fragments (known as \textit{reads}) of an organism’s genome generated by a sequencing machine against a well-studied reference genome.
The main goal of read mapping is to locate each read sequence in a reference genome, attempting to reassemble the reads back into their entire genome sequence. 
Read mapping remains one of the major performance bottlenecks in many genomic analyses for the three prominent sequencing technologies, Oxford Nanopore Technologies (ONT), PacBio HiFi, and Illumina~[\citenum{9923847, 10.1145/3503222.3507702}]. This is true even for the widely-used, well-maintained, state-of-the-art read mapper for modern CPUs, Minimap2~[\citenum{10.1093/bioinformatics/bty191}].

To understand the reasons behind read mapping's large performance overhead, we first briefly describe the workflow of Minimap2 in five key steps: 1) \textit{Index Construction}, 2) \textit{Seed Extraction}, 3) \textit{Index Querying}, 4) \textit{Anchor Sorting}, and 5) \textit{Seed Chaining}.
(1) \textit{Index Construction} step builds an index data structure to store short representative subsequences of length \emph{k} (known as seeds or \textit{k-mers}) along with their start location in the reference genome. 
(2) \textit{Seed Extraction} generates seeds from each read sequence using a similar algorithm to that of the \textit{Index Construction} step but without the need to store the generated seeds.
The small size of the seeds in the first two steps makes storing, retrieving, querying, and matching the seeds easier and more scalable compared to considering the complete genomic sequence.
(3) \textit{Index Querying} finds whether the seeds of the read exist in the reference genome by querying \and matching with the seeds stored in the index.
If they exist, then \textit{Index Querying} returns a set of matching pairs (known as anchors) containing the locations of the matching seed in both the read and the reference genome. 
(4) \textit{Anchor Sorting} sorts the retrieved anchors by their locations in the reference genome to narrow down the search space to the regions in the reference genome that share only the largest number of anchors (i.e., providing the highest similarity with a given read). 
(5) \textit{Seed Chaining} merges nearby matching anchors to build longer matching regions and exploits them to perform dynamic programming (DP) based sequence alignment~[\citenum{lindegger2023scrooge, 10.1093/bioinformatics/btaa777, 9251930}]. 

\begin{figure}[h]
\centering \includegraphics[width=0.6\textwidth]{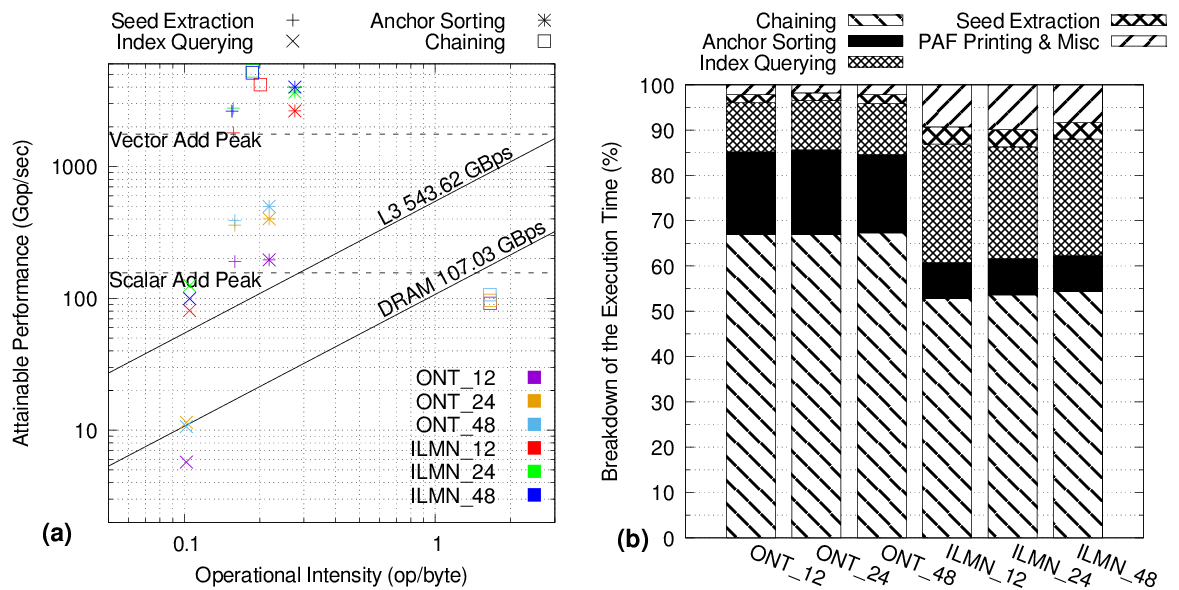}
     \caption{(a) Roofline model, and (b) execution time breakdown for the four key steps of the state-of-the-art read mapper, Minimap2, when mapping ONT and Illumina reads against the human reference genome (GRCh38). We use 12, 24, and 48 CPU threads and 2 Intel Xeon Gold 5118.
}\label{fig:profile}
\end{figure}

To systematically examine the contribution of each of the last four steps (\textit{Index Construction} is excluded as it is a one-time preprocessing step) to the total execution time of Minimap2, we calculate the roofline model~[\citenum{williams2009roofline}] and the execution time breakdown of Minimap2 in \figurename~\ref{fig:profile}, using Intel Advisor~[\citenum{8035181}].
We run Minimap2 in \textit{PAF mode} to exclude CIGAR calculation (i.e., sequence alignment) from the total execution time as 
1) We focus on improving the remaining computational steps since there is already a large body of work on proposing new algorithms and hardware accelerators for sequence alignment over the last five decades, 
2) Besides read mapping, many applications, such as genome assembly~[\citenum{vaser2017fast}] and building and mapping to pangenome reference~[\citenum{liao2023draft}], can benefit from fast CIGAR-less PAF calculations,
3) We can exploit the many existing works that provide fast sequence alignment, as we show in Table~\ref{tab:align}.


We use both real ONT (ultra-long) and Illumina (short) reads to evaluate different realistic behaviors of read mapping. 
We run Minimap2 using the default Illumina and ONT presets.
The roofline model plots the upper limit of achievable performance for executed code on a given processor against its operational intensity (operations count divided by memory traffic in bytes)~[\citenum{10.1145/1498765.1498785}]. It consists of horizontal peak performance rooflines and sloped memory bandwidth rooflines.


We make four key observations based on \figurename~\ref{fig:profile}. 
(1) {The \textit{Index Querying} step for both ONT and Illumina reads has a low operational intensity and a large amount of data movement for fetching \emph{all} locations of matching seeds from the index that is stored in main memory, i.e., dynamic random-access memory (DRAM).
The CPU can operate on the data only after the data is loaded from the off-chip main memory into CPU registers passing through cache hierarchy.
The low operational intensity and a large amount of data movement make} \textit{Index Querying} limited by the DRAM bandwidth, and increasing the number of CPU threads from 24 to 48 decreases attainable performance due to memory congestion.
The \textit{Index Querying} step includes complex irregular memory access patterns to fetch different index entries, leading to limited spatial locality, cache effectiveness, and frequent data movement between the memory subsystem and the processing units. 
(2) The attainable performance of both the \textit{Seed Extraction} and the \textit{Anchor Sorting} steps is above the L3 cache roofline regardless of the number of CPU threads. {Thus, both steps are not limited by the bandwidth of the DRAM {nor the bandwidth of the} L3 cache, and {they} have an access pattern that can be leveraged by the cache hierarchy in modern CPUs.}
(3) The \textit{Seed Chaining} step for ONT reads has the highest {operational} intensity compared to the three other steps{, while \textit{Anchor Sorting} for Illumina reads has the highest {operational} intensity}.
(4) Based on \figurename~\ref{fig:profile}(b), we observe that regardless of the number of {used CPU} threads, {the \textit{Seed Chaining}, \textit{Anchor Sorting}, and \textit{Index Querying} steps account for on average ~67\%, ~18\%, and ~11\% (54\%, 8\%, and 26\% using Illumina reads) of the total execution time using ONT reads, respectively}. 



We conclude that: (1) addressing the major compute-bound bottlenecks requires finding an alternative algorithmic method to {the largest contributor to the total execution time of ONT and Illumina read mapping,} \textit{Seed Chaining}, whose performance is mainly limited by the rate at which the CPU processes instructions, {and} (2) addressing the major memory-bound bottlenecks requires reducing back-and-forth data movements between the memory units and the processing units when running \textit{Index Querying}, whose performance {is the second largest contributor to the total execution time of Illumina read mapping and whose performance} is mainly limited by the rate at which data can be retrieved from the memory in both ONT and Illumina read mapping.

In this work, \textbf{our goal} is to significantly accelerate read mapping for both short and long reads by (1) introducing a computationally inexpensive algorithm to detect potential \ju{matching segment pairs} replacing the traditional computationally expensive \textit{Seed Chaining} step while maintaining comparable accuracy.
(2) overcoming the memory bottlenecks of the data-intensive index querying step by exploiting modern reconfigurable computing (i.e., FPGA) boards that tightly integrate high-bandwidth memory (HBM) with an FPGA chip in the same package{~[\citenum{10.1145/3491238}]}.
The integration of HBM with an FPGA allows higher memory density close to the computation fabric, an order of magnitude more memory bandwidth, and much lower latency to access stored data compared to traditional off-chip DRAM devices.
However, exploiting such modern FPGAs requires building efficient hardware architecture that handles the desired operations by leveraging only supported operations by the FPGA logic.
{Such modern FPGAs are already proven beneficial for sequence alignment}~[\citenum{10.1007/978-3-030-44534-8_2}] 
{and pre-alignment filtering}~[\citenum{9451578}]. 
{However, such new technology is not yet exploited for performing complete read mapping.}

To this end, we introduce \textit{\codename{}}, {the first} near-memory CPU-FPGA co-design for alleviating both the compute-bound and memory-bound bottlenecks in short and long-read mapping.
\codename{} is based on three \textbf{key ideas}:
(1) We observe that potential mapping locations {always} have the largest number of seed matches compared to other locations in the reference genome due to high similarity with a given read. 
\codename{} exploits this observation and {proposes} a {new} computational step with a linear time complexity in the number of seed matches that finds out the \ju{potential matching segment pairs} based on the {highest} number of seed matches scattered around a region in the reference genome. 
We call this approach \textit{Seed Voting}.
(2) \codename{} builds two new hardware architectures for performing \textit{Seed Extraction} and \textit{Index Querying} using modern FPGAs with HBM. 
Although \textit{Seed Extraction} is not memory-bound, it provides the input queries that are used for querying the index. 
Thus, minimizing the overall latency {requires accommodating} both steps, \textit{Seed Extraction} and \textit{Index Querying}, within the same FPGA chip.
(3) \codename{} {introduces} the first HBM-friendly hash table that is {specially designed to} exploit the access parallelism provided by modern FPGAs for \textit{fully} maximizing the querying throughput.
We carefully orchestrate execution on the CPU and the FPGA to hide data transfer latency and increase parallelism. 
\codename{} takes reads in FASTQ format and a reference genome (or precomputed index) in FASTA format and outputs mapping information in PAF format.


We summarize the \textbf{contributions} of this paper as follows:
\begin{itemize}
    \item We introduce \codename{}, the {first} software/hardware co-designed read mapper that exploits modern FPGAs featuring high bandwidth memory (HBM). {\codename{} is fully synthesizable, open-source, and ready-to-be-used on real hardware}.
    \item We provide, to our knowledge, the first FPGA accelerator for \textit{Seed Extraction} and \textit{Index Querying} for both short and long read mapping.
    \item We propose a new, efficient voting algorithm that replaces the compute-bound seed chaining algorithm while maintaining good accuracy.
    \item We experimentally demonstrate, using real ONT, HiFi, and Illumina sequences, that \codename{} outperforms Minimap2 by up to 40.3x, 4.8x, and 2.3x, respectively, when mapping the reads against the entire human reference genome.
    When performing read mapping with sequence alignment, \codename{} outperforms Minimap2 by 1.15-4.33 (using KSW2) and by 1.97-13.63x (using WFA-GPU).
\end{itemize}

%% file: paper/methods.tex
\section{Methods}\label{sec:methods}

\subsection{Overview}
\begin{figure}[h]
\centering 
\includegraphics[page=1,width=.6\textwidth]{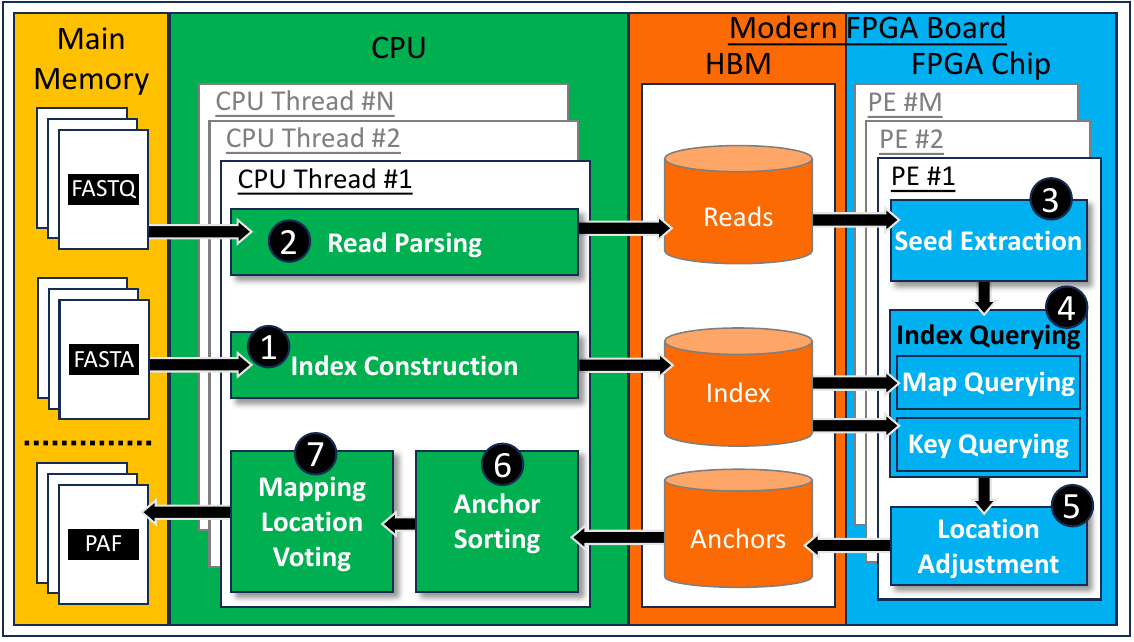}
     \caption{Overview of \codename{} that consists of a host CPU with main memory and a modern FPGA board that is equipped with an HBM memory.
}\label{fig:overview}
\end{figure}

\figurename~\ref{fig:overview} shows the overview of \codename{}, a CPU-FPGA co-design for accelerating read mapping. 
The pipeline can be divided into \ju{7} stages: \ju{\circled{1} \textit{Index Construction}}, \circled{2} \textit{Read Parsing}, \circled{3} \textit{Seed Extraction}, \circled{4} \textit{Index Querying}, \ju{\circled{5} \textit{Location Adjustment}}, 
\circled{6} \textit{Anchor Sorting}, and \circled{7} \textit{Mapping Location Voting}.
\sr{We explain each step in detail in the next subsections.}
Stages \circled{1}, \circled{2}, \circled{6}, and \circled{7} are performed on the host CPU as they better suit general-purpose CPUs and \ju{better benefit from CPU multithreading}.
Stages \circled{3}, \circled{4}, and \circled{5} are performed on FPGA featuring HBM as they better suit near-data FPGA acceleration.
\textbf{\codename{} efficiently uses both a host CPU and a modern FPGA in order to enable four different levels of parallelism}. 
First, the host CPU and the FPGA kernels are working concurrently. The host CPU launches the FPGA kernels asynchronously, such that the host CPU continues executing 
\ju{other stages} of \codename{} (i.e., \circled{2}, \circled{6}, \ju{and} \circled{7}) without the need to wait for the FPGAs. 
Second, \codename{} exploits the CPU multithreading for faster execution. 
\sr{\codename{} allocates the available CPU threads (e.g., \sr{a user-defined parameter,} $N$) and efficiently manages the tasks assigned to each CPU thread via thread-pool design pattern\ju{~[\citenum{10.5555/558986}]}.
Our thread-pool software design does not limit each CPU thread to processing a single read.
It rather keeps each CPU thread busy with any remaining task for any available read.
This achieves high allocation efficiency and optimized concurrent execution}.
\sr{The CPU threads are orchestrated such that the following five different tasks are quickly applied to each read sequence:} (1) Parsing \ju{the read sequences of a given FASTQ file using stage \circled{2}}, (2) transferring the parsed read \ju{sequence in batches} from the host CPU to the FPGA, (3) launching an FPGA kernel that executes stages \circled{3}, \circled{4}, and \circled{5}, (4) transferring the calculated anchors 
from the FPGA to the CPU, and (5) sorting the anchors using stage \circled{6}, \ju{performing \textit{Mapping Location Voting} using stage \circled{7}}, and writing the mapping results in PAF format. 
Third, by carefully building an efficient hardware architecture as a Processing Element (PE) for performing stages \circled{3}, \circled{4}, and \circled{5} on an FPGA chip, \codename{} is able to run multiple \ju{($M$)} PEs concurrently on the same FPGA chip for a higher level of parallelism. 
\ju{Fourth, \codename{} executes in dataflow manner~[\citenum{vitis}] where the PEs perform different tasks on an FPGA in parallel by} allowing consumer tasks to operate before the producer tasks have been completed. We describe in \sr{Section}~\ref{dataflow} the FPGA dataflow in more detail. 

\subsection{HBM Organization}

\begin{figure}[t]
\centering \includegraphics[width=0.6\textwidth]{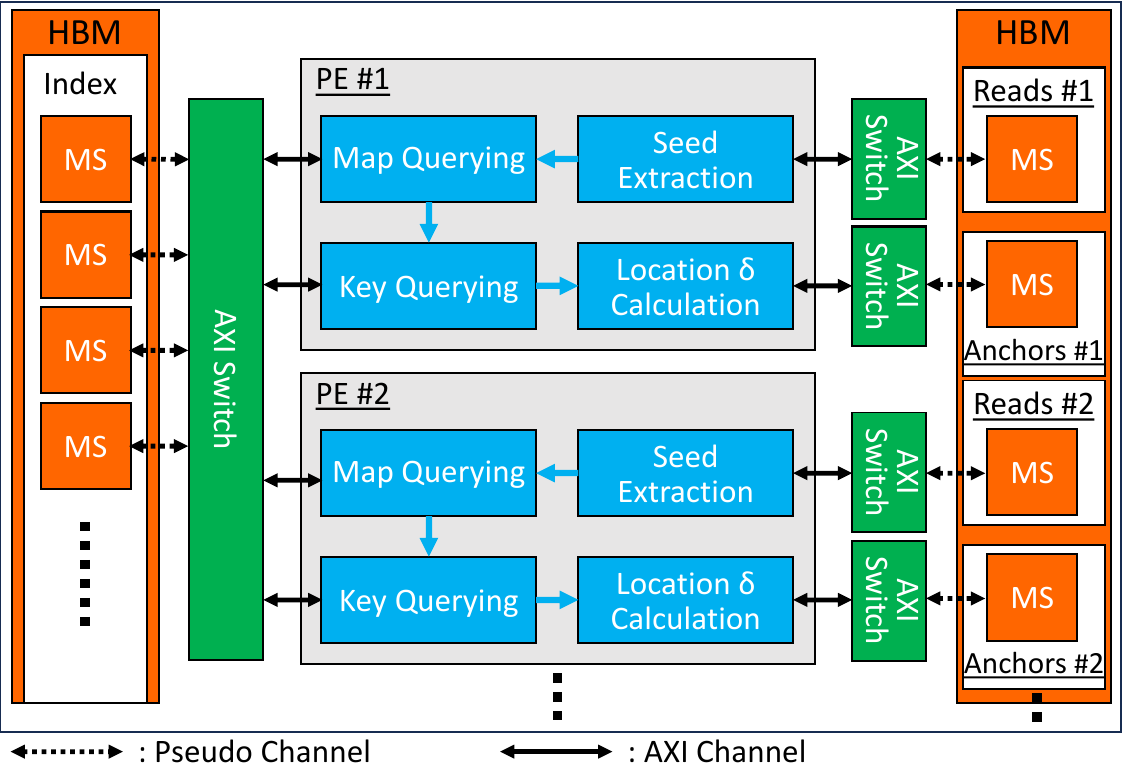}
     \caption{Near-memory FPGA design of \codename{}}\label{fig:fpga}
\end{figure}

To mitigate the memory bottleneck caused by the data transfer between the memory and the computing elements, modern FPGA features HBM.
\figurename~\ref{fig:fpga} depicts the internal organization of an HBM that consists of two main components: 1) HBM stacks and 2) an HBM controller inside the FPGA. A stack comprises multiple memory sections (MSs), each of which is connected to the HBM controller through a 64-bit pseudo channel. 
In the HBM controller, each pseudo channel is connected to an AXI channel that interacts with the user logic. 
Each AXI channel is associated with a pseudo channel and can directly access the aligned MS. 
To make each AXI channel able to access the full HBM space (i.e., all the MSs), an AXI switch is integrated between the AXI channels and the pseudo channels. 
However, to reach the maximum bandwidth and the minimum latency for the HBM controller, direct routing from the AXI port to the aligned pseudo channel should be used, and accessing unaligned MS should be avoided~[\citenum{9114755}]. 
As a result, to optimize the throughput of our design, we 1) partition our data into batches with a smaller size than the size of MS \sr{(e.g., Xilinx Alveo U55C features two 8GB HBM2 memories, each of which has 16 512MB MSs)}, and 2) \ju{\sr{carefully design the architecture of each PE such that each AXI channel can access only} a unique MS, i.e., limiting the size of the memory space accessed by each AXI channel to the size of one MS.} 

\subsection{Index Processing}

\subsubsection{Index Construction}
The purpose of the index is to efficiently store \sr{extracted information (e.g., \emph{seeds} and their start locations in the reference genome) from the subject reference genome and efficiently facilitate the querying and retrieval of such information when needed}. 
For a given reference genome and a set of parameters \sr{(e.g., seed length)}, the index only needs to be built once and can be reused for different sets of read sequences. 
\sr{As building the index is not considered as a contributor to the total execution time of read mapping}, we build the index at the host CPU. 
\sr{\codename{} uses} the minimizer algorithm~[\citenum{10.1093/bioinformatics/bth408}] to \sr{choose the seeds to be stored in the reference genome index}. 
The minimizer algorithm uses a \textit{hash and compare} heuristic: (1) It computes the hash value of \sr{\textit{w}} consecutive\sr{/overlapping} k-mers (a subsequence of length \textit{k}), and (2) compares all the hash \sr{values} and outputs \sr{the k-mer with} the smallest hash value as a \sr{resulting} minimizer seed that represents the subject k-mers. 
\ju{The \textit{Index Construction} step of \codename{} is fully configurable for different \textit{w} and \textit{k} values.
The implementation is multi-threaded, and its execution time has the same order of magnitude as Minimap2.} 

We build an index data structure that is similar to the \textit{HashMap}~[\citenum{10.5555/1404505}] as it offers (1) high data locality \sr{and (2) constant-time performance for adding or querying any seed}.
\sr{The high data locality leads to} \ju{a higher throughput. The data are accessed into contiguous blocks, which leverage the memory architecture and enable multiple hardware accelerations such as burst transfer.} 
\sr{The constant-time performance leads to a constant number of required clock cycles for performing index-dependent operations and a constant number of memory accesses for fetching indexed data.
This helps in easily orchestrating the index querying step with all other steps that depend on its output and thus increasing task-level parallelism.
This index data structure has two arrays: a map array and a key array.}
The map array stores pointers to the key array and is indexed by the seed value (i.e., a hash value of a seed).
The key array stores the locations of extracted seeds in the reference genome.

Some seeds can occur very frequently in the reference genome and, as a result, can increase the rate of false-positive mapping locations and unnecessarily increase the time spent to query the index and process the seed locations~[\citenum{10.1093/bioinformatics/btv670}].
To overcome this issue, we remove from the index the seeds (along with their locations) that occur (based on each seed's number of locations) more frequently than \sr{a user-defined} value of \texttt{max\_occ}. 


\subsubsection{Index Storing}
\sr{As frequent accesses to the index stored in the main memory cause memory bottlenecks, \codename{}} stores the index directly in the HBM of the FPGA.
\sr{This provides two key advantages: (1) minimizes data communication latency due to shorter interconnects between the FPGA chip and HBM compared to the interconnects between the CPU and the main memory, and (2) provides an order of magnitude more bandwidth than traditional main memory (e.g., DDR4).
Since the size of each MS in HBM is limited to a fixed size and the size of the index depends on the subject reference genome, we partition both the map and key arrays of the index into subarrays, each of which has a size smaller than or equal to the size of an MS.
By storing each subarray in a different MS, \codename{} can handle any index of any size as long as the sizes of the index, one batch of read sequences, and one batch of anchors collectively do not exceed the HBM capacity} (e.g., \ju{16GB on the Xilinx Alveo U55C}).
The index is loaded in the HBM of the FPGA \emph{only} once before the execution of the read mapper.

\subsubsection{Index Querying}
The purpose of the \textit{Index Querying} stage is to efficiently retrieve all \sr{occurrence} locations in the reference genome for a \sr{given query} seed.
To maximize the throughput of this stage: (1) we minimize the number of memory accesses, and (2) our design only accesses consecutive memory addresses to leverage burst transfers. 

Our \textit{Index Querying} is a two-step mechanism\sr{: accessing the map array and accessing the key array of the index}. Both steps perform unique memory access \sr{(i.e., unique entry in the arrays)} for each seed, and both steps are performed in parallel.
The first step is to access the map array with the value of the seed, which returns \sr{two pointers (i.e., addresses to the corresponding memory section) that indicate the start and the end of the list of seed locations stored in the key array.} 
The second step is to fetch all the locations between the start and end entries in the key array. \ju{Each fetched location from the index (corresponding to a location in the reference genome) is then associated with the corresponding location of seed in the read to form an anchor.}
\sr{To perform} both steps in parallel (through pipelining), each PE is connected to the index through 2 different AXI channels. The first AXI channel is used to access the map array, and the second one to access the key array.

\subsection{Read Processing}
\subsubsection{Read Parsing \& Storing}
The goal of \sr{\textit{Read Parsing \& Storing}} is to convert the input read sequences stored as \sr{FASTQ} files into sequences that can efficiently be stored in the HBM and processed by the FPGA logic. 
To efficiently overlap FPGA processing time with data transfer time and minimize the HBM allocation size for accommodating read sequences, the reads are transferred and processed in batches. 
We construct the read batches at the CPU side and transfer them to the HBM of the FPGA. 
To maximize the bandwidth between the FPGA logic and the HBM, we limit the size of \sr{each read} batch to the size of an MS of the HBM. Since our FPGA design only performs \textit{Seed Extraction} and \textit{Index Querying}, there is no need to store any metadata \ju{(read ID, read len, read number)} on the HBM. 
Each read batch consists of a \sr{stream} of read sequences concatenated to each other, where each read sequence is separated by a special character \texttt{E} \sr{(a different character from the read alphabets, A, C, G, T, and N)}. The metadata \ju{(e.g., read ID, read length, number of reads)} for a given batch of reads is stored \sr{at} the CPU side.

Each \sr{read} batch is \sr{transferred} \ju{by a CPU thread} to the HBM of the FPGA through the PCIe interface. The FPGA can process as many batches as the number of PEs implemented in the FPGA chip in parallel. Therefore we limit the number of batches to the number of PEs\ju{, which we discuss in Section~\ref{dataflow}}. 

\subsubsection{Seed Extraction}\label{seed_extraction}
The goal of the \textit{Seed Extraction} is to quickly extract the seeds of each read stored in the batches. \sr{Similar to the \textit{Index Construction} step of \codename{}, \codename also uses} the minimizer algorithm~[\citenum{10.1093/bioinformatics/bth408}] to \sr{extract the seeds from read sequences}. 
Our hardware \sr{architecture} of \textit{Seed Extraction} step calculates {one} minimizer seed every cycle. 
To reach this performance, we use {two} key approaches: (1) We replicate the hardware logic responsible for computing the hash values \sr{\textit{w} times}, and thus it allows us to compute the hash value of the \sr{subject \textit{w}} consecutive k-mers in parallel. (2) Our implementation is pipelined\sr{, which means the critical path delay of each PE is shortened by dividing it into stages or smaller tasks.
This allows \codename{} to meet target timing constraints (e.g., maximum operating frequency) 
and achieve more parallelism by calcullating multiple minimizer seeds in parallel}.

\subsection{Calculating the Mapping Locations}
\subsubsection{Anchor Sorting}
The goal of the \textit{Anchor Sorting} stage is to sort the anchors according to their location in the reference genome. Sorting the anchors allows us to quickly identify the \ju{potential matching segment pairs} during the voting stage. 
Based on the literature~\ju{\citenum{Mueller2012}} and our experimental evaluation, there is no FPGA implementation for sorting algorithms that is faster than multicore CPU implementations as used in Minimap2~[\citenum{10.1093/bioinformatics/bty191}].
\ju{Our FPGA implementation of a pseudo-in-place merge sort algorithm shows one order of magnitude higher execution time compared to 24-threads CPU implementation}.
For this reason, we decide to perform \textit{Anchor Sorting} \sr{at} the CPU side and not on \sr{the} FPGA \sr{chip}.
We implement two types of sorting algorithms: radix sort and merge sort~\ju{\citenum{Knuth1973}}. 
We observe that for Illumina reads, merge sort is 1.84x faster than radix sort, while for ONT reads, radix sort is 1.32x faster than merge sort.

\subsubsection{Location Adjustment and Mapping Location Voting}
The goal of the \textit{Mapping Location Voting} stage is to \sr{quickly find the potential \ju{matching segment pairs between a given read sequence and the reference genome}}.
The key idea of \textit{Voting} \sr{is based on the observation that the} correct mapping location \ju{always} \sr{has the largest number} of anchors compared to the other mapping locations due to the high similarity between the read sequence and the sequence extracted at the correct mapping location in the reference genome.
Based on this observation, we develop a linear time (in the number of anchors) voting algorithm.

Our voting mechanism consists of two main steps. The first one is performed on the FPGA \ju{after} \textit{Index Querying}, and it consists of subtracting the location of the seed within the read sequence from the location of the seed within the reference genome. \ju{The list of subtracted locations ($\delta$) along with the corresponding location within the read sequence, also called the list of anchors (\textit{A}) constitutes the input of the second step.} 
The second step is the core of our algorithm, and it is performed on the CPU after \textit{Anchor Sorting}. During this step, we iterate once through the list of sorted anchors, and based on those, we output a list of \ju{matching segment pairs that have the highest number of votes.}

Our voting mechanism is different from the one used in Genome-on-Diet~[\citenum{alser2022genome}] in two different aspects. 
(1) \codename{} only performs one round of voting on the whole read to identify all subsequences in the read that share a large number of votes with the reference genome.
The goal of \codename{} is to only identify the correct mapping locations in the reference genome for each of these subsequences and report them in the PAF file.
Genome-on-Diet on the other hand performs multiple rounds of voting on multiple subsequences of the read to map one or more of the read subsequences (two subsequences with a large gap in between) together to cover structural variations (SVs) occurring in the read.
The linked subsequences are needed to generate a CIGAR string that represents the SV.
(2) The index data structure, the \textit{Seed Extraction} algorithm, and the indexing parameters that \codename{} uses are \emph{all} different than the one used in Genome-on-Diet. 
\codename{} uses minimizer seeds, while Genome-on-Diet uses sparsified seeds that span a much larger region in the reference genome compared to that of \codename{}.
Thus, \codename{} also uses a different implementation and different parameters for our voting algorithm than the one used in Genome-on-Diet.

To explain our voting algorithm, let a list of anchors be $A$, and the $i^{th}$ and $j^{th}$  anchors \sr{are represented} as pairs of integer numbers $(L^i_{read}, L^i_{ref})$ and $(L^j_{read}, L^j_{ref})$), respectively.
While $L^i_{read}$ and $L^j_{read}$ represent the locations of different seeds within the same read, $L^i_{ref}$ and $L^j_{ref}$ represent the locations of these seeds within the reference genome. 
Let $e^i_j$ be the total number of deletions and insertions between \sr{the $i^{th}$ and $j^{th}$ anchors, such that we} have the following inequality:
$$| (L^j_{read} - L^i_{read}) - (L^j_{ref} -  L^i_{ref}) | \le e^i_j$$

This inequality becomes equality if there are only deletions or insertions between the two seed matches.
\sr{Let the} subtracted locations for the two anchors \sr{be}: $\delta^i = L^i_{ref} - L^i_{read}$ and  $\delta^j = L^j_{ref} - L^j_{read}$, \sr{such that} the following inequality holds:
$$| \delta^j -  \delta^i| \le e^i_j$$

\noindent Thus the difference between two subtracted locations $\Delta^i_j = | \delta^j -  \delta^i|$ gives us a lower bound for the total number of insertions and deletions between two anchors. If there are only insertions or deletions, then $\Delta^i_j = e^i_j$.
For anchors that are close to each other, we expect $\Delta^i_j$ to be \ju{close to} $e^i_j$ since the number of consecutive insertions and deletions is small. 

\textbf{Location Adjustment}. Thereby, it makes sense to sort the list of locations based on $\delta$ and then iterate through the list. 
For this reason, the goal of \textit{Location Adjustment} step is to compute the $\delta$ \sr{values} on the FPGA chip after performing \textit{Index Querying} step.
\ju{On FPGA, the computation of the $\delta$ values is performed in parallel with the other stages performed on the FPGA and thus has no cost in terms of execution time.}

To match segments from the read sequence to segments from the reference genome, we define a voting distance \texttt{vt\_dist}. \ju{We consider that two anchors $i$ and $j$ belong to the same segment if the total number of insertions and deletions between the two anchors is smaller than the user-defined voting distance. (i.e., $e^i_j \le \texttt{vt\_dist}$). Computing the exact value of $e^i_j$ is computationally expensive and requires DP (it requires performing alignment between the anchors). Since for anchors that are close to each other $\Delta^i_j$ can be seen as a good approximation of $e^i_j$, and the $\Delta^i_j$ of all the consecutive anchors can be computed with a linear time complexity, we consider that the two anchors belong to the same segment if $\Delta^i_j \le \texttt{vt\_dist}$.} 

The voting distance can be arbitrarily large depending on the application. If we want to perform alignment on the output segments, the voting distance should have the same order of magnitude as the alignment bandwidth. \ju{Indeed, for a given segment pair, we might have a  $\Delta^i_j$ such that $\Delta^i_j = \texttt{vt\_dist}$. Now if we consider that we only have insertions or deletions between the two anchors, the following holds $\Delta^i_j = e^i_j$ and thus $e^i_j =\texttt{vt\_dist}$. In order to align a segment pair having a section with only \texttt{vt\_dist} deletions or insertions, we need a bandwidth of at least  \texttt{vt\_dist}.}
For each matching segment pair, we define a voting score corresponding to the number of anchors belonging to the given segment. We use the voting score as a metric to measure the quality of the matching segment pair. The higher the voting score is, the more anchors belong to the segment pair and the higher the probability of \sr{being the} correct mapping location.

Our voting algorithm (Algorithm~\ref{alg:voting}) takes the list of sorted anchors as input and outputs a list of matching segment pairs with the highest voting score that meet some user-defined constraints, such as the minimum length of the segments. The algorithm starts by initializing two temporary mutable segment pairs, \ju{one corresponding to the positive strand and the other to the negative}. 
The voting algorithm then iterates over the list of sorted anchors. For each iteration, we check if the anchor belongs to the temporary segment pair of the corresponding strand (Line~\ref{alg:comp}). If yes, we adjust the boundaries of the segment pair based on the anchor and increment the voting score (Line~\ref{alg:update}). Else, we check if the voting score of the temporary segment pair is greater than the lowest voting score of the list of mapping segment pairs and if the temporary segment pair meets the user-defined constraints. If both conditions are met, we append the segment pair to the list of matching segment pairs (and remove the segment pair with the lowest voting score if necessary) (Line~\ref{alg:append}). We then initialize the temporary segment pair with the current anchor  (Line~\ref{alg:init_with}).

%% file: paper/results.tex
\section{Results}

We evaluate 1) the time for data transfer and processing per genomic base (bp), 2) the FPGA resource utilization, 3) the end-to-end speedup of \codename{} compared to Minimap2, and 4) the accuracy of \codename{} compared to Minimap2. We provide all the commands used to generate the results and a comprehensive treatment of all evaluation results on the \codename{} GitHub page. Our evaluated datasets and presets are provided in the Supplementary Materials, Sections \ref{datasetsec} and \ref{datapresetsec}.
We implement our accelerator designs on an Alveo U55C card featuring the Xilinx Virtex Ultrascale+ XCU55C with 16 GiB HBM2 connected to an AMD EPYC 7302P host system with 64 GiB of memory. All the experiments, including the CPU-only experiments, are run on the described system.

\subsection{Data Transfer and Processing Time Analysis}\label{perf}
\begin{figure}[h]
         \centering \includegraphics[width=0.6\textwidth]{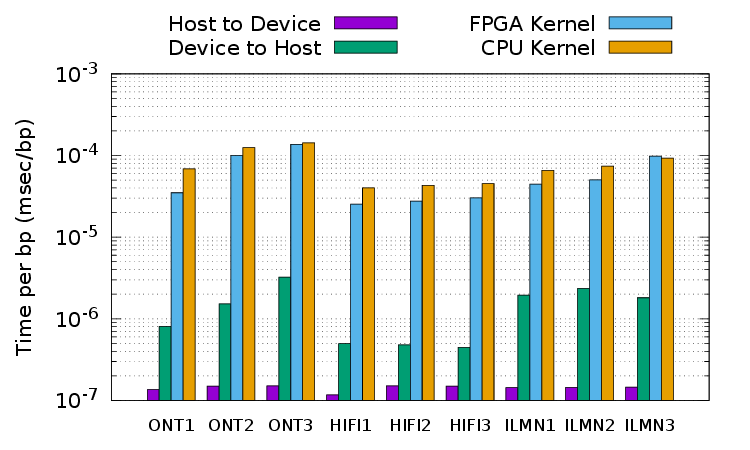}
     \caption{Data transfer time and processing time per bp, when transferring the reads to the FPGA board, transferring the mapping locations back to the host CPU, executing the FPGA kernel, and executing a CPU implementation of the seeding kernel on the host CPU.}
     \label{fig:time}
\end{figure}

In \figurename~\ref{fig:time}, we evaluate the data transfer time from the host CPU to the FPGA board and from the FPGA board back to the host CPU, the FPGA kernel processing time, and the processing time of a CPU-optimized version of the seeding kernel running on the host CPU. 
We use 32 CPU threads for the CPU-based step and 8 FPGA PEs for the FPGA-based step.
We perform our measurements when running the complete pipeline of \codename{}. 
We use the nine presets and normalize the time to the number of bases by dividing the transfer time or processing time by the batch size. 
Based on \figurename~\ref{fig:time} we make three key observations. (1) Our FPGA kernel is always faster than the CPU version except for the ILMN3 preset and provides up to 1.96x, 1.58x and 1.47x speedup for ONT, HiFi, and Illumina reads, respectively, compared to the CPU kernel. 
The highest performance compared to the CPU kernel is reached when using small \texttt{max\_occ} values. 
This is expected, as for large \texttt{max\_occ} values, the average number of locations returned by the key table is larger. 
Since the returned locations are stored contiguously in the memory, the CPU cache hierarchy and the data prefetching mechanisms can be leveraged by the CPU to increase the overall throughput of the CPU kernel. (2) Increasing \texttt{max\_occ} value always increases the execution time of our FPGA kernel. This is expected as increasing \texttt{max\_occ} increases the number of locations to fetch from the HBM. (3) The transfer time is always more than 20x faster than the FPGA kernel execution time. Transferring the locations from the device to the host is always slower than transferring the read sequences from the host to the device. Since we use asynchronous programming in our host program, we trigger the host-to-device and the device-to-host transfers in parallel. Thus, we are only limited by the device-to-host transfer time.

We conclude that outsourcing the entire seeding stage on FPGA is always beneficial since it reduces the CPU workload, the processing time on FPGA is always faster or comparable to the CPU processing time, and the transfer time is negligible compared to the processing time.  

\subsection{FPGA Resource Utilization}
We list the resource utilization of \codename{} on the FPGA in \tablename~\ref{tab:resources}. For each sequencing technology and from the FPGA design point-of-view, the three presets we choose differ only in the values of \texttt{k} and \texttt{k} as \texttt{max\_occ} and \texttt{vt\_dist} impact only the CPU-based steps and batch size impacts the memory allocation.
Thus, we report the resource utilization for each sequencing technology and not for each preset.
From \tablename~\ref{tab:resources}, we observe that regardless of the sequencing technology, there are always enough resources to theoretically accommodate more than 16 PEs.
However, in practice, we are limited by the number of HBM AXI Channels. Since each PE is designed to use 4 AXI Channels, the maximum number of PEs that \codename{} can accommodate is 8 to cope with the 32 memory channels offered by the HBM of the board we are using.

\begin{table}[h]
\caption{FPGA resource utilization (in \%) for different sequencing data types (i.e., \texttt{w} and \texttt{k} values) and different numbers of PEs.}\label{tab:resources}
\begin{tabular*}{\columnwidth}{@{\extracolsep{\fill}}cccccc}
\toprule
 &  \textbf{PEs} & \textbf{CLB} & \textbf{LUT} & \textbf{FF} & \textbf{BRAM} \\
\midrule
\multirow{2}*{ONT} & 1 & 18.88 & 10.33 & 7.05 & 10.69\\

            & 8 & 31.62 & 18.22 & 13.49 & 16.07\\
\hline
\multirow{2}*{HiFi} & 1 & 19.30 & 10.51 & 7.21 & 10.69\\

            & 8 & 33.89 & 19.67 & 14.72 & 16.07\\
\hline
\multirow{2}*{Illumina} & 1 & 19.32 & 10.51 & 7.17 & 10.69\\

            & 8 & 33.38 & 19.66 & 14.41 & 16.07\\
\botrule
\end{tabular*}
\end{table}

\subsection{End-to-End Speedup}\label{speed}

\begin{figure}[h]
         \centering \includegraphics[width=0.6\textwidth]{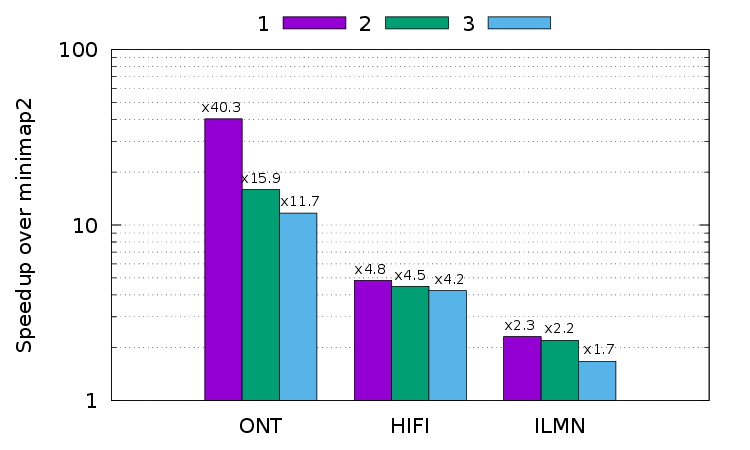}
     \caption{End-to-end speedup of \codename{} over Minimap2 for the nine different presets. We run Minimap2 without performing sequence alignment.}
     \label{fig:speed}
\end{figure}

We evaluate the end-to-end speedup that is provided by \codename{} over Minimap2. 
As a baseline, we run Minimap2 without alignment using 32 CPU threads on the same host CPU used by \codename{} for a fair comparison.
We build the index used by \codename{} and Minimap2 beforehand so that the execution time for \textit{Index Construction} steps is not accounted for in the total execution time. 
We run \codename{} using 32 CPU threads and 8 FPGA PEs, as we discussed in the previous subsection. 
We load the read sequences and the index into DRAM before each run of \codename{} and Minimap2 to reduce the impact of the I/O costs. \figurename~\ref{fig:speed} presents the speedup of \codename{} over Minimap2 for the nine presets.

We make three key observations. (1) \codename{} provides the largest speedup rate when using ONT reads. This is expected for two main reasons: 1) ONT preset uses small k and w values, which causes the number of locations returned by the index to be large, and 2) the length of evaluated ONT reads is much larger (between 10k and 100k bps) than that of the evaluated HiFi and Illumina reads. 
Consequently, the number of extracted seeds, the number of queried seeds, and the number of returned locations per read are much larger than those for HiFi and Illumina reads.
The larger workload benefits directly from FPGA acceleration and high parallelism offered by \codename{}.
Minimap2 uses chaining, which has a quadratic time complexity in the number of returned locations, and the voting algorithm used in \codename{} has a linear time complexity in the number of locations. 
So for small k and w values and ultra-long reads, \textit{Mapping Location Voting} provides a non-negligible speedup compared to chaining. 
(2) Using a small \texttt{max\_occ} as in \texttt{ONT1} leads to have the highest speedup rate (40.3$\times$). This is expected as it reduces the number of returned locations after querying the index and hence there is a smaller workload to be sorted and performing voting step on, which reduces the overall execution time.
(3) For large values of k (HiFi and Illumina presets), the impact of \texttt{max\_occ} is less important on the end-to-end speedup. Indeed using large values for k increases the number of unique minimizers and decreases the average occurrence of each minimizer. Thus increasing \texttt{max\_occ} while having a small average occurrence only has a limited impact on the execution time.
\ju{Whereas for ONT (k = 15)} \texttt{max\_occ} \ju{has a large impact on the end-to-end speedup. For ONT, the execution time when using} \texttt{max\_occ} \ju{= 10 is 2.4x faster than the execution time when using} \texttt{max\_occ} = 50.

We conclude that, in terms of speedup, \codename{} performs the best for long and inaccurate reads compared to Minimap2. We also conclude that, in terms of execution time, the choice of the \texttt{max\_occ} value is impactful for long and inaccurate reads.

\subsection{Accuracy Analysis}\label{accuracy}
\begin{figure}[h]
\centering \includegraphics[width=0.6\textwidth]{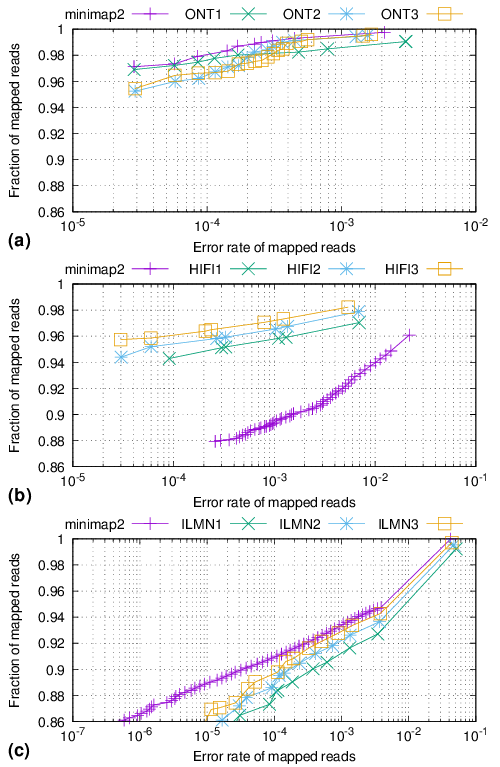}
     \caption{Read mapping accuracy of \codename{} compared to Minimap2, using mapeval from PAFtools. 
     }\label{fig:accuracy}
\end{figure}

We evaluate the accuracy of \codename{} compared to Minimap2 using simulated human reads and using mapeval tool from the PAFtools library provided by Minimap2. 
Simulated reads were mapped to the complete Human reference genome GRCh38. A read is considered correctly mapped if its longest mapping overlaps with the true interval, and the overlap length is $\geq$10\% of the true interval length. We run Minimap2 using its default presets for each read type.
We measure the accuracy of \codename{} for the nine presets. We provide in \figurename~\ref{fig:accuracy} the \textit{(error rate, fraction of mapped reads)} pairs that are above different mapping quality thresholds.

Based on \figurename~\ref{fig:accuracy}, we make three key observations. (1) For accurate read sequences (HiFi and Illumina), increasing \texttt{max\_occ} always increases the accuracy. 
This is not true for long noisy reads (ONT) accuracy results. A possible explanation is that increasing \texttt{max\_occ} also increases the rate of false positive seed matches (i.e., random seed matches due to, for example, highly repetitive seeds in Human data). Since the amount of false positive seed matches is higher for noisy reads, it also leads to a higher number of false positive votes. (2) For HiFi reads, \codename{} has always a better accuracy even with \texttt{max\_occ} set to 1 compared to Minimap2. 
For ONT and Illumina reads, \codename{} has always a lower fraction (<2\%) of mapped reads for the same error rate compared to Minimap2 
(3) For HiFi and Illumina, we observe that the accuracy converges to an upper bound. Choosing a \texttt{max\_occ} value above 5 and 450 for HiFi and Illumina, respectively, only has a limited effect on the fraction of mapped reads.

We conclude that even if \codename{} uses a lightweight pre-alignment filtering algorithm, \textit{Mapping Location Voting}, compared to chaining, \codename{} provides high accuracy compared to Minimap2 for all sequencing data types. 

\subsection{Performing Sequence Alignment}

We examine in Table~\ref{tab:align} the benefits of integrating the existing state-of-the-art sequence aligners with \codename{} to perform complete read mapping with sequence alignment.
We choose one representative tool from each of the four directions for accelerating sequence alignment: 
1) Using modern processors that provide wider registers (e.g., 512-bit wide) for executing vector operations on multiple operands at once for high parallelism. We choose a recent, fastest vectorized implementation~[\citenum{Kalikar2022}] of the widely-used aligner, KSW2~[\citenum{10.1093/bioinformatics/btp324}]. It accelerates KSW2 by up to $2.2\times$. We refer to this recent implementation in Table~\ref{tab:align} as \textit{KSW2 AVX}.
2) Building CMOS-based customized hardware architectures to speed up the alignment process. We choose a non-optimal alignment algorithm, called GACT~[\citenum{turakhia2018darwin}], that has such an accelerator. It divides the DP matrix into overlapping submatrices and greedily processing each submatrix using systolic arrays. We refer to it in Table~\ref{tab:align} as \textit{GACT CMOS}. 
3) Exploiting a large number of threads and large local memory provided by modern GPUs to compute alignments of many independent sequence pairs concurrently. We choose a recent GPU implementation~[\citenum{Aguado-Puig2022.04.18.488374}] of the wavefront algorithm (WFA)~[\citenum{marco2021fast}], which reformulates the classic Smith-Waterman-Gotoh recursion and shows significant speedups for highly similar sequence pairs. 
The GPU implementation~[\citenum{Aguado-Puig2022.04.18.488374}] of the WFA algorithm improves the original CPU implementation by 1.5-7.7$\times$ using long reads.
We refer to it in Table~\ref{tab:align} as \textit{WFA GPU}. 
4) Using a pre-alignment filtering algorithm to reduce the number of mapping locations to be verified by sequence alignment by providing approximate edit distance calculation. We choose SneakySnake~[\citenum{10.1093/bioinformatics/btaa1015}] as representative since it provides the highest accuracy and speedup compared to other algorithms~[\citenum{ALSER20224579}].
We refer to it in Table~\ref{tab:align} as \textit{SneakySnake CPU}. 

We present in Table~\ref{tab:align} the read mapping throughput of Minimap2 (which uses KSW2) and \codename{} integrated with each of the representative tools that we discuss. 
We observe that integrating existing tools for sequence alignment with \codename{} is always beneficial. 
It provides up to $13.63\times$, $13.67\times$, and $3.89\times$ higher read mapping throughput compared to Minimap2.



\begin{table}[h]
\caption{Read mapping throughput (number of mapped reads per second) of Minimap2 and \codename{} integrated with state-of-the-art pre-alignment filter and sequence alignment tools}\label{tab:align}
\begin{tabular}{cccccc}
\hline
\multicolumn{1}{l}{} & \multicolumn{4}{c}{\textbf{\hspace{40 mm}\codename{} integrated with\hspace{40 mm} }}               & \multicolumn{1}{l}{\multirow{2}{*}{\textbf{\hspace{6 mm}Minimap2\hspace{6 mm}}}} \\ \cline{2-5}
                     & \textbf{KSW2 AVX} & \textbf{GACT CMOS} & \textbf{WFA GPU} & \textbf{SneakySnake CPU} & \multicolumn{1}{l}{}                                                      \\ \hline
\multicolumn{1}{c|}{\textbf{ONT}} &
  \begin{tabular}[c]{@{}c@{}}1'516 (4.33x)\end{tabular} &
  \begin{tabular}[c]{@{}c@{}}3'037 (8.67x)\end{tabular} &
  \textbf{\begin{tabular}[c]{@{}c@{}}4'771 (13.63x)\end{tabular}} &
  \begin{tabular}[c]{@{}c@{}}1'324 (3.78x)\end{tabular} &
  350 \\ \hline
\multicolumn{1}{c|}{\textbf{HiFi}} &
  \begin{tabular}[c]{@{}c@{}}1'287  (1.15x)\end{tabular} &
  \begin{tabular}[c]{@{}c@{}}2'752 (2.47x)\end{tabular} &
  \textbf{\begin{tabular}[c]{@{}c@{}}15'237 (13.67x)\end{tabular}} &
  \begin{tabular}[c]{@{}c@{}}4'774 (4.28x)\end{tabular} &
  1'114 \\ \hline
\multicolumn{1}{c|}{\textbf{Illumina}} &
  \begin{tabular}[c]{@{}c@{}}281'827 (2.44x)\end{tabular} &
  \begin{tabular}[c]{@{}c@{}}156'168 (1.35x)\end{tabular} &
  \begin{tabular}[c]{@{}c@{}}228'205 (1.97x)\end{tabular} &
  \textbf{\begin{tabular}[c]{@{}c@{}}450'062 (3.89x)\end{tabular}} &
  115'511 \\ \hline
\end{tabular}
\end{table}




%% file: paper/discussion.tex
\vspace{-0.5cm}

\section{Conclusion}\label{sec:discussion}

We demonstrate that we can use the HBM of modern FPGAs to mitigate the memory bottleneck of the index querying step. We propose an index data structure that leverages the HBM organization. We develop an FPGA+HBM design that performs the seed extraction and index querying steps. We implement our design on a real FPGA with 8 PEs.
In addition, we propose a lightweight voting algorithm with a linear time complexity that replaces the computationally expensive chaining step while maintaining good accuracy.
We integrate our FPGA design and our voting algorithm into a CPU-FPGA co-designed read mapping tool, \codename{}. We experimentally demonstrate, using real ONT, HiFi, and Illumina sequences, that \codename{} outperforms Minimap2 by up to 40.3x, 4.8x, and 2.3x, respectively, when mapping the reads against the entire human reference genome.

%% file: paper/funding.tex
\section*{Funding}

\sr{We acknowledge the generous gifts of our industrial partners, including Intel and VMware. This work is also partially supported by the European Union's Horizon programme for research and innovation [101047160 - BioPIM] and the Swiss National Science Foundation (SNSF) $[200021\_213084]$.}

%% file: paper/appendix.tex
\onecolumn
 \renewcommand{\thesection}{S\arabic{section}}
 \section{Supplementary Materials}
\setcounter{table}{0}
\setcounter{figure}{0}
\renewcommand{\thetable}{Supplementary \arabic{table}}
\renewcommand\thefigure{Supplementary \arabic{figure}}

\renewcommand\photoname{Supplementary Table}

\begin{algorithm}
\caption{Voting Algorithm}\label{alg:voting}
\begin{algorithmic}[1]
\Require List of sorted anchors $A$
\Ensure List of matching segment pairs $SP$
\State $sp \gets [sp\_init(), sp\_init()]$ \Comment{Initialize the temporary segment pairs for both strands}\label{alg:init}
\For{$a$ \textbf{in} $A$}
\State $str \gets a.strand$
\If{$a.\delta \le sp.\delta[str] + vt\_dist$} \label{alg:comp}
\State $sp[str].update(a)$ \Comment{Adjust the segment pair boundaries and increment the voting score}\label{alg:update}
\Else
\State $SP.append\_if(sp[str])$ \label{alg:append}
\State $sp[str] \gets sp\_init\_with(a)$ \label{alg:init_with}
\EndIf
\EndFor
\State $SP.append\_if(sp[0])$
\State $SP.append\_if(sp[1])$
\end{algorithmic}
\end{algorithm}

\subsection{FPGA Dataflow}\label{dataflow}

We use a high-level synthesis (HLS) design flow~[\citenum{vitis}] to implement and map our accelerator design. We present in \figurename~\ref{fig:fpga} the four key tasks performed on the FPGA by each PE. (1) We design the first minimizer-based hardware implementation to extract the seeds from the read sequences stored in the HBM. Our hardware module computes one seed every cycle as described in Subsection~\ref{seed_extraction}.
(2) We use the seeds computed in the previous stage to access the map array stored in the HBM and retrieve the start and end positions of the list of locations of the given seed within the key array. This task is pipelined with an interval of 2 since we need to access the map array twice to get the start and end positions of the list of locations. 
(3) We read the consecutive locations between the start and end positions computed at the previous stage in the key arrays stored in the HBM and concatenate them with the location of the seed within the read sequence to form anchors that we send to the next task. The number of cycles for this task highly depends on the number of locations in the index for the input seed. 
(4) \ju{We compute the Location $\delta$ by subtracting the location of the seed within the read sequence from the location of the seed within the reference genome.}

\ju{To avoid complex synchronization primitives, each PE has its own read sequence and anchor buffer. The index buffer is shared among all the PEs since accessing the index doesn't require synchronization between the PEs.}

We use the DATAFLOW pragma to enable task-level pipelining. In addition, we use FIFOs, known as \texttt{hls::stream}, to connect the different dataflow tasks allowing the producer and the consumer tasks to work in parallel.

\subsection{Evaluated Datasets}
\label{datasetsec}
We use the complete human reference genome GRCh38 to perform all our experiments. To evaluate the throughput of our FPGA design (Subsection~\ref{perf}) and the end-to-end speedup of \codename{} (Subsection~\ref{speed}), we choose three real sequencing read sets for Ashkenazi Son HG002 (NA24385) provided by the Genome in a Bottle (GIAB) Project~[\citenum{zook2016extensive}].
The three sequencing read sets represent the current state-of-the-art and prominent sequencing technologies, ultra-long reads from ONT, accurate long HiFi reads from PacBio, and short reads from Illumina. \textbf{Supplementary Table}~\ref{tab:reads} provides the details of these datasets. To evaluate the accuracy of \codename{} (Subsection~\ref{accuracy}), we simulate ONT and HiFi reads using PBSIM3~[\citenum{10.1093/nargab/lqac092}] with the error models of \texttt{ERRHMM-ONT.model} and \texttt{ERRHMM-SEQUEL.model}, respectively. We simulate Illumina reads with mason2~[\citenum{fu_mi_publications962}] with option \texttt{-illumina-prob-mismatch-scale 2.5}.

\subsection{Evaluated Presets}
\label{datapresetsec}

In \textbf{Supplementary Table}~\ref{tab:presets}, we provide different presets that we use to run \codename{}. 
For \texttt{k} and \texttt{w}, we choose the same default values as those of Minimap2. 
For each type of sequencing technology, we empirically select three different \texttt{max\_occ} parameters to evaluate different \textit{speed vs accuracy} tradeoffs. 
For a given \texttt{max\_occ}, we select a \texttt{vt\_dist} that provides high accuracy. To select the batch size, we first find the upper bound that does not lead to overflow of the memory sections (512 MB) allocated for the output, and we then select the batch size that optimizes the overall speedup. 
For all the presets, our FPGA accelerator operates at 317MHz.

\begin{photo*}[h]
\label{app}
\caption{Details of real sequencing read sets used in read mapping evaluation. We calculate the statistics using NanoPlot tool.}\label{tab:reads}
\begin{tabular*}{\textwidth}{c|ccc}
\toprule
\textbf{Read set} & \textbf{Short reads (Illumina)} & \textbf{Accurate, long reads (HiFi)} & \textbf{Ultra-long reads (ONT)} \\
\midrule
Direct download link & 
\href{https://ftp-trace.ncbi.nlm.nih.gov/ReferenceSamples/giab/data/AshkenazimTrio/HG002_NA24385_son/NIST_Illumina_2x250bps/reads/D1_S1_L001_R1_001.fastq.gz}{{D1\_S1\_L001\_R1\_001.fastq}} & \href{https://ftp-trace.ncbi.nlm.nih.gov/ReferenceSamples/giab/data/AshkenazimTrio/HG002_NA24385_son/PacBio_CCS_15kb_20kb_chemistry2/reads/m64011_190830_220126.fastq.gz}{{m64011\_190830\_220126.fastq}} & \href{https://ftp-trace.ncbi.nlm.nih.gov/ReferenceSamples/giab/data/AshkenazimTrio/HG002_NA24385_son/Ultralong_OxfordNanopore/guppy-V3.4.5/HG002_ONT-UL_GIAB_20200204.fastq.gz}{\small{HG002\_ONT-UL\_GIAB\_20200204.fastq}}$^{\mathrm{a}}$ \\
Mean read length & 248.2 & 18,491.2 & 15,806.4 \\
Mean read quality & 8.8 & 31.3 & 9.2 \\
Number of reads & 122,495,089 & 1,423,276 & 2,000,000 \\
Read length N50 & 250 & 18,563 & 49,093 \\
STDEV read length & 9.3 &  2,184.5 & 34,618.1 \\
Total bases & 30,405,193,199 & 26,318,110,120 & 31,612,763,195 \\
Longest read length & 250 & 46,910 & 1,331,423 \\
\botrule
\multicolumn{4}{l}{\small{$^{\mathrm{a}}$We consider only the first 2 million 
reads whose length is greater than or equal 1000 bp (using \texttt{NanoFilt --length 1000})}}
\end{tabular*}
\end{photo*}

\begin{photo*}[h]
\caption{Seeding and Indexing Presets}\label{tab:presets}
\begin{tabular*}{\textwidth}{@{\extracolsep{\fill}}c|ccc|ccc|ccc}
\toprule
 &  \multicolumn{3}{c|}{\textbf{ONT (k = 15, w = 10)}} &\multicolumn{3}{c|}{\textbf{HiFi (k = 19, w = 19)}} & \multicolumn{3}{c}{\textbf{Illumina (k = 21, w = 11)}} \\
 \midrule
 Preset & ONT1 & ONT2 & ONT3 & HIFI1 & HIFI2 & HIFI3 & ILMN1 & ILMN2 & ILMN3 \\
\texttt{max\_occ} & 10 & 50 & 100 & 1 & 2 & 5 & 50 & 150 & 450\\
\texttt{vt\_dist} & 950 & 750 & 700 & 5000 & 4000 & 4000 & 10 & 10 & 10 \\
batch size & 64Mbps & 32MBbps & 16MBbps & \multicolumn{3}{c|}{128Mbps} & \multicolumn{3}{c}{32Mbps}\\

\botrule
\end{tabular*}
\end{photo*}


\newpage
\subsection{Related FPGA accelerators for Read Mapping}
To our knowledge, \codename{} is the first CPU-FPGA co-designed read mapper that overcomes the compute and memory bottlenecks for both short and long reads. There exists a large body of work on accelerating read mapping using field programmable logic but in contrast to \codename{}, only a few of them propose an end-to-end integration.

\textbf{FPGA Acceleration of Pre-alignment Filtering.}  Some prior works accelerate the pre-alignment filtering step that aims to quickly eliminate dissimilar sequences before invoking computationally-expensive alignment algorithms, using FPGA~[\citenum{ 10.1093/bioinformatics/btz234,10.1093/bioinformatics/btaa1015, 9451578,10.1093/bioinformatics/btx342, alser2017magnet}] and other hardware acceleration platforms~[\citenum{kim2018grim, 9251930, bingol2021gatekeeper}]. These works are orthogonal to our work, and they can be integrated with a complete read mapping pipeline (e.g., \codename{}) to provide, for example, approximate edit distance calculation. 

\textbf{FPGA Acceleration of Sequence Alignment.} In recent years, a number of FPGA accelerators have been proposed to improve the performance of sequence alignment~[\citenum{Fei2018, HAGHI202339, ALSER20224579}]. Many other works target modern CPUs~[\citenum{Kalikar2022}], GPUs~[\citenum{Ahmed2019, Aguado-Puig2022.04.18.488374,lindegger2022algorithmic,lindegger2023scrooge}], and even emerging processing-in-memory~[\citenum{diab2023framework,mao2022genpip}] to accelerate the sequence alignment step.
These acceleration works can be exploited to complement \codename{} for providing fast read mapping with sequence alignment.

\textbf{End-to-End Read Mapper on (SoC-)FPGA.} 
[\citenum{9520260}] propose an energy-efficient read mapper on SoC-FPGA based on the q-gram lemma. However, the proposed work was not tested on the entire human reference genome but only on single chromosomes, and the execution time of the proposed work is slower than state-of-the-art CPU read mappers. [\citenum{9320533}] propose an FPGA accelerator for short read mapping based on the seed-and-extend algorithm. However, the proposed work is only compared to older FPGA read mappers and not to the state-of-the-art CPU mappers for short reads~[\citenum{10.1093/bioinformatics/btp324, 8820962}]. Unlike our work, none of these works are open-sourced, and we are unable to experimentally evaluate them.

%% file: main.bbl
\begin{thebibliography}{52}
\providecommand{\natexlab}[1]{#1}
\providecommand{\url}[1]{\texttt{#1}}
\expandafter\ifx\csname urlstyle\endcsname\relax
  \providecommand{\doi}[1]{doi: #1}\else
  \providecommand{\doi}{doi: \begingroup \urlstyle{rm}\Url}\fi

\bibitem[Alser et~al.(2020{\natexlab{a}})Alser, Bingöl, Cali, Kim, Ghose,
  Alkan, and Mutlu]{9154510}
Mohammed Alser, Zülal Bingöl, Damla~Senol Cali, Jeremie Kim, Saugata Ghose,
  Can Alkan, and Onur Mutlu.
\newblock {Accelerating Genome Analysis: A Primer on an Ongoing Journey}.
\newblock \emph{IEEE Micro}, 40\penalty0 (5):\penalty0 65--75,
  2020{\natexlab{a}}.
\newblock \doi{10.1109/MM.2020.3013728}.

\bibitem[Alser et~al.(2022{\natexlab{a}})Alser, Lindegger, Firtina, Almadhoun,
  Mao, Singh, Gomez-Luna, and Mutlu]{ALSER20224579}
Mohammed Alser, Joel Lindegger, Can Firtina, Nour Almadhoun, Haiyu Mao,
  Gagandeep Singh, Juan Gomez-Luna, and Onur Mutlu.
\newblock {From molecules to genomic variations: Accelerating genome analysis
  via intelligent algorithms and architectures}.
\newblock \emph{CSBJ}, 20:\penalty0 4579--4599, 2022{\natexlab{a}}.
\newblock ISSN 2001-0370.
\newblock \doi{https://doi.org/10.1016/j.csbj.2022.08.019}.
\newblock URL
  \url{https://www.sciencedirect.com/science/article/pii/S2001037022003531}.

\bibitem[Vaser et~al.(2017)Vaser, Sovi{\'c}, Nagarajan, and
  {\v{S}}iki{\'c}]{vaser2017fast}
Robert Vaser, Ivan Sovi{\'c}, Niranjan Nagarajan, and Mile {\v{S}}iki{\'c}.
\newblock {Fast and accurate de novo genome assembly from long uncorrected
  reads}.
\newblock \emph{Genome research}, 27\penalty0 (5):\penalty0 737--746, 2017.

\bibitem[Liao et~al.(2023)Liao, Asri, Ebler, Doerr, Haukness, Hickey, Lu,
  Lucas, Monlong, Abel, et~al.]{liao2023draft}
Wen-Wei Liao, Mobin Asri, Jana Ebler, Daniel Doerr, Marina Haukness, Glenn
  Hickey, Shuangjia Lu, Julian~K Lucas, Jean Monlong, Haley~J Abel, et~al.
\newblock {A draft human pangenome reference}.
\newblock \emph{Nature}, 617\penalty0 (7960):\penalty0 312--324, 2023.

\bibitem[Alser et~al.(2021)Alser, Rotman, Deshpande, Taraszka, Shi, Baykal,
  Yang, Xue, Knyazev, Singer, Balliu, Koslicki, Skums, Zelikovsky, Alkan,
  Mutlu, and Mangul]{Alser2021}
Mohammed Alser, Jeremy Rotman, Dhrithi Deshpande, Kodi Taraszka, Huwenbo Shi,
  Pelin~Icer Baykal, Harry~Taegyun Yang, Victor Xue, Sergey Knyazev,
  Benjamin~D. Singer, Brunilda Balliu, David Koslicki, Pavel Skums, Alex
  Zelikovsky, Can Alkan, Onur Mutlu, and Serghei Mangul.
\newblock {Technology dictates algorithms: recent developments in read
  alignment}.
\newblock \emph{Genome Biology}, 22\penalty0 (1):\penalty0 249, Aug 2021.
\newblock ISSN 1474-760X.
\newblock \doi{10.1186/s13059-021-02443-7}.
\newblock URL \url{https://doi.org/10.1186/s13059-021-02443-7}.

\bibitem[LaPierre et~al.(2020)LaPierre, Alser, Eskin, Koslicki, and
  Mangul]{lapierre2020metalign}
Nathan LaPierre, Mohammed Alser, Eleazar Eskin, David Koslicki, and Serghei
  Mangul.
\newblock {Metalign: efficient alignment-based metagenomic profiling via
  containment min hash}.
\newblock \emph{Genome biology}, 21\penalty0 (1):\penalty0 1--15, 2020.

\bibitem[Meyer et~al.(2022)Meyer, Fritz, Deng, Koslicki, Lesker, Gurevich,
  Robertson, Alser, Antipov, Beghini, et~al.]{meyer2022critical}
Fernando Meyer, Adrian Fritz, Zhi-Luo Deng, David Koslicki, Till~Robin Lesker,
  Alexey Gurevich, Gary Robertson, Mohammed Alser, Dmitry Antipov, Francesco
  Beghini, et~al.
\newblock {Critical assessment of metagenome interpretation: the second round
  of challenges}.
\newblock \emph{Nature methods}, 19\penalty0 (4):\penalty0 429--440, 2022.

\bibitem[LaPierre et~al.(2019)LaPierre, Mangul, Alser, Mandric, Wu, Koslicki,
  and Eskin]{lapierre2019micop}
Nathan LaPierre, Serghei Mangul, Mohammed Alser, Igor Mandric, Nicholas~C Wu,
  David Koslicki, and Eleazar Eskin.
\newblock {MiCoP: microbial community profiling method for detecting viral and
  fungal organisms in metagenomic samples}.
\newblock \emph{BMC genomics}, 20:\penalty0 1--10, 2019.

\bibitem[Mao et~al.(2022{\natexlab{a}})Mao, Alser, Sadrosadati, Firtina,
  Baranwal, Senol~Cali, Manglik, Alserr, and Mutlu]{9923847}
Haiyu Mao, Mohammed Alser, Mohammad Sadrosadati, Can Firtina, Akanksha
  Baranwal, Damla Senol~Cali, Aditya Manglik, Nour~Almadhoun Alserr, and Onur
  Mutlu.
\newblock {GenPIP: In-Memory Acceleration of Genome Analysis via Tight
  Integration of Basecalling and Read Mapping}.
\newblock In \emph{2022 55th IEEE/ACM International Symposium on
  Microarchitecture (MICRO)}, pages 710--726, 2022{\natexlab{a}}.
\newblock \doi{10.1109/MICRO56248.2022.00056}.

\bibitem[Mansouri~Ghiasi et~al.(2022)Mansouri~Ghiasi, Park, Mustafa, Kim,
  Olgun, Gollwitzer, Senol~Cali, Firtina, Mao, Almadhoun~Alserr,
  Ausavarungnirun, Vijaykumar, Alser, and Mutlu]{10.1145/3503222.3507702}
Nika Mansouri~Ghiasi, Jisung Park, Harun Mustafa, Jeremie Kim, Ataberk Olgun,
  Arvid Gollwitzer, Damla Senol~Cali, Can Firtina, Haiyu Mao, Nour
  Almadhoun~Alserr, Rachata Ausavarungnirun, Nandita Vijaykumar, Mohammed
  Alser, and Onur Mutlu.
\newblock {GenStore: A High-Performance in-Storage Processing System for Genome
  Sequence Analysis}.
\newblock In \emph{Proceedings of the 27th ACM International Conference on
  Architectural Support for Programming Languages and Operating Systems},
  ASPLOS '22, page 635–654, New York, NY, USA, 2022. Association for
  Computing Machinery.
\newblock ISBN 9781450392051.
\newblock \doi{10.1145/3503222.3507702}.
\newblock URL \url{https://doi.org/10.1145/3503222.3507702}.

\bibitem[Li(2018)]{10.1093/bioinformatics/bty191}
Heng Li.
\newblock {Minimap2: pairwise alignment for nucleotide sequences}.
\newblock \emph{Bioinformatics}, 34\penalty0 (18):\penalty0 3094--3100, 05
  2018.
\newblock ISSN 1367-4803.
\newblock \doi{10.1093/bioinformatics/bty191}.
\newblock URL \url{https://doi.org/10.1093/bioinformatics/bty191}.

\bibitem[Lindegger et~al.(2023)Lindegger, Senol~Cali, Alser, G{\'o}mez-Luna,
  Mansouri~Ghiasi, and Mutlu]{lindegger2023scrooge}
Jo{\"e}l Lindegger, Damla Senol~Cali, Mohammed Alser, Juan G{\'o}mez-Luna, Nika
  Mansouri~Ghiasi, and Onur Mutlu.
\newblock {Scrooge: A Fast and Memory-Frugal Genomic Sequence Aligner for CPUs,
  GPUs, and ASICs}.
\newblock \emph{Bioinformatics}, 2023.

\bibitem[Marco-Sola et~al.(2020)Marco-Sola, Moure, Moreto, and
  Espinosa]{10.1093/bioinformatics/btaa777}
Santiago Marco-Sola, Juan~Carlos Moure, Miquel Moreto, and Antonio Espinosa.
\newblock {Fast gap-affine pairwise alignment using the wavefront algorithm}.
\newblock \emph{Bioinformatics}, 37\penalty0 (4):\penalty0 456--463, 09 2020.
\newblock ISSN 1367-4803.
\newblock \doi{10.1093/bioinformatics/btaa777}.
\newblock URL \url{https://doi.org/10.1093/bioinformatics/btaa777}.

\bibitem[Senol~Cali et~al.(2020)Senol~Cali, Kalsi, Bingöl, Firtina,
  Subramanian, Kim, Ausavarungnirun, Alser, Gomez-Luna, Boroumand, Norion,
  Scibisz, Subramoneyon, Alkan, Ghose, and Mutlu]{9251930}
Damla Senol~Cali, Gurpreet~S. Kalsi, Zülal Bingöl, Can Firtina, Lavanya
  Subramanian, Jeremie~S. Kim, Rachata Ausavarungnirun, Mohammed Alser, Juan
  Gomez-Luna, Amirali Boroumand, Anant Norion, Allison Scibisz, Sreenivas
  Subramoneyon, Can Alkan, Saugata Ghose, and Onur Mutlu.
\newblock {GenASM: A High-Performance, Low-Power Approximate String Matching
  Acceleration Framework for Genome Sequence Analysis}.
\newblock In \emph{2020 53rd Annual IEEE/ACM International Symposium on
  Microarchitecture (MICRO)}, pages 951--966, 2020.
\newblock \doi{10.1109/MICRO50266.2020.00081}.

\bibitem[Williams et~al.(2009{\natexlab{a}})Williams, Waterman, and
  Patterson]{williams2009roofline}
Samuel Williams, Andrew Waterman, and David Patterson.
\newblock {Roofline: an insightful visual performance model for multicore
  architectures}.
\newblock \emph{Communications of the ACM}, 52\penalty0 (4):\penalty0 65--76,
  2009{\natexlab{a}}.

\bibitem[Marques et~al.(2017)Marques, Duarte, Ilic, Sousa, Belenov, Thierry,
  and Matveev]{8035181}
Diogo Marques, Helder Duarte, Aleksandar Ilic, Leonel Sousa, Roman Belenov,
  Philippe Thierry, and Zakhar~A. Matveev.
\newblock {Performance Analysis with Cache-Aware Roofline Model in Intel
  Advisor}.
\newblock In \emph{2017 International Conference on High Performance Computing
  \& Simulation (HPCS)}, pages 898--907, 2017.
\newblock \doi{10.1109/HPCS.2017.150}.

\bibitem[Williams et~al.(2009{\natexlab{b}})Williams, Waterman, and
  Patterson]{10.1145/1498765.1498785}
Samuel Williams, Andrew Waterman, and David Patterson.
\newblock {Roofline: An Insightful Visual Performance Model for Multicore
  Architectures}.
\newblock \emph{Commun. ACM}, 52\penalty0 (4):\penalty0 65–76, apr
  2009{\natexlab{b}}.
\newblock ISSN 0001-0782.
\newblock \doi{10.1145/1498765.1498785}.
\newblock URL \url{https://doi.org/10.1145/1498765.1498785}.

\bibitem[Shi et~al.(2022)Shi, Kara, Hagleitner, Diamantopoulos, Syrivelis, and
  Alonso]{10.1145/3491238}
Runbin Shi, Kaan Kara, Christoph Hagleitner, Dionysios Diamantopoulos, Dimitris
  Syrivelis, and Gustavo Alonso.
\newblock {Exploiting HBM on FPGAs for Data Processing}.
\newblock \emph{ACM Trans. Reconfigurable Technol. Syst.}, 15\penalty0 (4), dec
  2022.
\newblock ISSN 1936-7406.
\newblock \doi{10.1145/3491238}.
\newblock URL \url{https://doi.org/10.1145/3491238}.

\bibitem[Ansaloni et~al.(2020)Ansaloni, Scarabottolo, and
  Pozzi]{10.1007/978-3-030-44534-8_2}
Giovanni Ansaloni, Ilaria Scarabottolo, and Laura Pozzi.
\newblock {Judiciously Spreading Approximation Among Arithmetic Components with
  Top-Down Inexact Hardware Design}.
\newblock In Fernando Rinc{\'o}n, Jes{\'u}s Barba, Hayden K.~H. So, Pedro
  Diniz, and Juli{\'a}n Caba, editors, \emph{Applied Reconfigurable Computing.
  Architectures, Tools, and Applications}, pages 14--29, Cham, 2020. Springer
  International Publishing.
\newblock ISBN 978-3-030-44534-8.

\bibitem[Singh et~al.(2021)Singh, Alser, Cali, Diamantopoulos, Gómez-Luna,
  Corporaal, and Mutlu]{9451578}
Gagandeep Singh, Mohammed Alser, Damla~Senol Cali, Dionysios Diamantopoulos,
  Juan Gómez-Luna, Henk Corporaal, and Onur Mutlu.
\newblock {FPGA-Based Near-Memory Acceleration of Modern Data-Intensive
  Applications}.
\newblock \emph{IEEE Micro}, 41\penalty0 (4):\penalty0 39--48, 2021.
\newblock \doi{10.1109/MM.2021.3088396}.

\bibitem[Garg and Sharapov(2001)]{10.5555/558986}
Rajat~P. Garg and Illya Sharapov.
\newblock \emph{{Techniques for Optimizing Applications: High Performance
  Computing}}.
\newblock Prentice Hall PTR, USA, 2001.
\newblock ISBN 0130934763.

\bibitem[Xilinx(2022)]{vitis}
Xilinx.
\newblock {Vitis High-Level Synthesis User Guide (UG1399)}, jun 2022.
\newblock URL \url{https://docs.xilinx.com/r/2022.1-English/ug1399-vitis-hls}.

\bibitem[Wang et~al.(2020)Wang, Huang, Zhang, and Alonso]{9114755}
Zeke Wang, Hongjing Huang, Jie Zhang, and Gustavo Alonso.
\newblock {Shuhai: Benchmarking High Bandwidth Memory On FPGAs}.
\newblock In \emph{2020 IEEE 28th FCCM}, pages 111--119, 2020.
\newblock \doi{10.1109/FCCM48280.2020.00024}.

\bibitem[Roberts et~al.(2004)Roberts, Hayes, Hunt, Mount, and
  Yorke]{10.1093/bioinformatics/bth408}
Michael Roberts, Wayne Hayes, Brian~R. Hunt, Stephen~M. Mount, and James~A.
  Yorke.
\newblock {Reducing storage requirements for biological sequence comparison}.
\newblock \emph{Bioinformatics}, 20\penalty0 (18):\penalty0 3363--3369, 07
  2004.
\newblock ISSN 1367-4803.
\newblock \doi{10.1093/bioinformatics/bth408}.
\newblock URL \url{https://doi.org/10.1093/bioinformatics/bth408}.

\bibitem[Mehlhorn and Sanders(2008)]{10.5555/1404505}
Kurt Mehlhorn and Peter Sanders.
\newblock \emph{{Algorithms and Data Structures: The Basic Toolbox}}.
\newblock Springer Publishing Company, Incorporated, 1 edition, 2008.
\newblock ISBN 9783540779773.

\bibitem[Xin et~al.(2015)Xin, Nahar, Zhu, Emmons, Pekhimenko, Kingsford, Alkan,
  and Mutlu]{10.1093/bioinformatics/btv670}
Hongyi Xin, Sunny Nahar, Richard Zhu, John Emmons, Gennady Pekhimenko, Carl
  Kingsford, Can Alkan, and Onur Mutlu.
\newblock {Optimal seed solver: optimizing seed selection in read mapping}.
\newblock \emph{Bioinformatics}, 32\penalty0 (11):\penalty0 1632--1642, 11
  2015.
\newblock ISSN 1367-4803.
\newblock \doi{10.1093/bioinformatics/btv670}.
\newblock URL \url{https://doi.org/10.1093/bioinformatics/btv670}.

\bibitem[Mueller et~al.(2012)Mueller, Teubner, and Alonso]{Mueller2012}
Rene Mueller, Jens Teubner, and Gustavo Alonso.
\newblock {Sorting networks on FPGAs}.
\newblock \emph{The VLDB Journal}, 21\penalty0 (1):\penalty0 1--23, Feb 2012.
\newblock ISSN 0949-877X.
\newblock \doi{10.1007/s00778-011-0232-z}.
\newblock URL \url{https://doi.org/10.1007/s00778-011-0232-z}.

\bibitem[Knuth(1973)]{Knuth1973}
Donald Knuth.
\newblock \emph{The Art Of Computer Programming, vol. 3: Sorting And
  Searching}.
\newblock Addison-Wesley, 1973.

\bibitem[Alser et~al.(2022{\natexlab{b}})Alser, Eudine, and
  Mutlu]{alser2022genome}
Mohammed Alser, Julien Eudine, and Onur Mutlu.
\newblock {Genome-on-Diet: Taming Large-Scale Genomic Analyses via Sparsified
  Genomics}.
\newblock \emph{arXiv}, 2022{\natexlab{b}}.

\bibitem[Kalikar et~al.(2022)Kalikar, Jain, Vasimuddin, and Misra]{Kalikar2022}
Saurabh Kalikar, Chirag Jain, Md~Vasimuddin, and Sanchit Misra.
\newblock {Accelerating minimap2 for long-read sequencing applications on
  modern CPUs}.
\newblock \emph{Nature Computational Science}, 2\penalty0 (2):\penalty0 78--83,
  Feb 2022.
\newblock ISSN 2662-8457.
\newblock \doi{10.1038/s43588-022-00201-8}.
\newblock URL \url{https://doi.org/10.1038/s43588-022-00201-8}.

\bibitem[Li and Durbin(2009)]{10.1093/bioinformatics/btp324}
Heng Li and Richard Durbin.
\newblock {Fast and accurate short read alignment with Burrows–Wheeler
  transform}.
\newblock \emph{Bioinformatics}, 25\penalty0 (14):\penalty0 1754--1760, 05
  2009.
\newblock ISSN 1367-4803.
\newblock \doi{10.1093/bioinformatics/btp324}.
\newblock URL \url{https://doi.org/10.1093/bioinformatics/btp324}.

\bibitem[Turakhia et~al.(2018)Turakhia, Bejerano, and
  Dally]{turakhia2018darwin}
Yatish Turakhia, Gill Bejerano, and William~J Dally.
\newblock {Darwin: A genomics co-processor provides up to 15,000 x acceleration
  on long read assembly}.
\newblock \emph{ACM SIGPLAN Notices}, 53\penalty0 (2):\penalty0 199--213, 2018.

\bibitem[Aguado-Puig et~al.(2022)Aguado-Puig, Marco-Sola, Moure, Matzoros,
  Castells-Rufas, Espinosa, and Moreto]{Aguado-Puig2022.04.18.488374}
Quim Aguado-Puig, Santiago Marco-Sola, Juan~Carlos Moure, Christos Matzoros,
  David Castells-Rufas, Antonio Espinosa, and Miquel Moreto.
\newblock {WFA-GPU: Gap-affine pairwise alignment using GPUs}.
\newblock \emph{bioRxiv}, 2022.
\newblock \doi{10.1101/2022.04.18.488374}.
\newblock URL
  \url{https://www.biorxiv.org/content/early/2022/04/18/2022.04.18.488374}.

\bibitem[Marco-Sola et~al.(2021)Marco-Sola, Moure, Moreto, and
  Espinosa]{marco2021fast}
Santiago Marco-Sola, Juan~Carlos Moure, Miquel Moreto, and Antonio Espinosa.
\newblock {Fast gap-affine pairwise alignment using the wavefront algorithm}.
\newblock \emph{Bioinformatics}, 37\penalty0 (4):\penalty0 456--463, 2021.

\bibitem[Alser et~al.(2020{\natexlab{b}})Alser, Shahroodi, Gómez-Luna, Alkan,
  and Mutlu]{10.1093/bioinformatics/btaa1015}
Mohammed Alser, Taha Shahroodi, Juan Gómez-Luna, Can Alkan, and Onur Mutlu.
\newblock {{SneakySnake: a fast and accurate universal genome pre-alignment
  filter for CPUs, GPUs and FPGAs}}.
\newblock \emph{Bioinformatics}, 36\penalty0 (22-23):\penalty0 5282--5290, 12
  2020{\natexlab{b}}.
\newblock ISSN 1367-4803.
\newblock \doi{10.1093/bioinformatics/btaa1015}.
\newblock URL \url{https://doi.org/10.1093/bioinformatics/btaa1015}.

\bibitem[Zook et~al.(2016)Zook, Catoe, McDaniel, Vang, Spies, Sidow, Weng, Liu,
  Mason, Alexander, et~al.]{zook2016extensive}
Justin~M Zook, David Catoe, Jennifer McDaniel, Lindsay Vang, Noah Spies, Arend
  Sidow, Ziming Weng, Yuling Liu, Christopher~E Mason, Noah Alexander, et~al.
\newblock Extensive sequencing of seven human genomes to characterize benchmark
  reference materials.
\newblock \emph{Scientific data}, 3\penalty0 (1):\penalty0 1--26, 2016.

\bibitem[Ono et~al.(2022)Ono, Hamada, and Asai]{10.1093/nargab/lqac092}
Yukiteru Ono, Michiaki Hamada, and Kiyoshi Asai.
\newblock {PBSIM3: a simulator for all types of PacBio and ONT long reads}.
\newblock \emph{NAR Genomics and Bioinformatics}, 4\penalty0 (4), 12 2022.
\newblock ISSN 2631-9268.
\newblock \doi{10.1093/nargab/lqac092}.
\newblock URL \url{https://doi.org/10.1093/nargab/lqac092}.
\newblock lqac092.

\bibitem[Holtgrewe(2010)]{fu_mi_publications962}
M.~Holtgrewe.
\newblock {Mason - A Read Simulator for Second Generation Sequencing Data}.
\newblock \emph{Technical Report FU Berlin}, October 2010.
\newblock URL \url{http://publications.imp.fu-berlin.de/962/}.

\bibitem[Alser et~al.(2019)Alser, Hassan, Kumar, Mutlu, and
  Alkan]{10.1093/bioinformatics/btz234}
Mohammed Alser, Hasan Hassan, Akash Kumar, Onur Mutlu, and Can Alkan.
\newblock {Shouji: a fast and efficient pre-alignment filter for sequence
  alignment}.
\newblock \emph{Bioinformatics}, 35\penalty0 (21):\penalty0 4255--4263, 03
  2019.
\newblock ISSN 1367-4803.
\newblock \doi{10.1093/bioinformatics/btz234}.
\newblock URL \url{https://doi.org/10.1093/bioinformatics/btz234}.

\bibitem[Alser et~al.(2017{\natexlab{a}})Alser, Hassan, Xin, Ergin, Mutlu, and
  Alkan]{10.1093/bioinformatics/btx342}
Mohammed Alser, Hasan Hassan, Hongyi Xin, Oğuz Ergin, Onur Mutlu, and Can
  Alkan.
\newblock {GateKeeper: a new hardware architecture for accelerating
  pre-alignment in DNA short read mapping}.
\newblock \emph{Bioinformatics}, 33\penalty0 (21):\penalty0 3355--3363, 05
  2017{\natexlab{a}}.
\newblock ISSN 1367-4803.
\newblock \doi{10.1093/bioinformatics/btx342}.
\newblock URL \url{https://doi.org/10.1093/bioinformatics/btx342}.

\bibitem[Alser et~al.(2017{\natexlab{b}})Alser, Mutlu, and
  Alkan]{alser2017magnet}
Mohammed Alser, Onur Mutlu, and Can Alkan.
\newblock {MAGNET: understanding and improving the accuracy of genome
  pre-alignment filtering}.
\newblock \emph{arXiv preprint arXiv:1707.01631}, 2017{\natexlab{b}}.

\bibitem[Kim et~al.(2018)Kim, Senol~Cali, Xin, Lee, Ghose, Alser, Hassan,
  Ergin, Alkan, and Mutlu]{kim2018grim}
Jeremie~S Kim, Damla Senol~Cali, Hongyi Xin, Donghyuk Lee, Saugata Ghose,
  Mohammed Alser, Hasan Hassan, Oguz Ergin, Can Alkan, and Onur Mutlu.
\newblock {GRIM-Filter: Fast seed location filtering in DNA read mapping using
  processing-in-memory technologies}.
\newblock \emph{BMC genomics}, 19\penalty0 (2):\penalty0 23--40, 2018.

\bibitem[Bing{\"o}l et~al.(2021)Bing{\"o}l, Alser, Mutlu, Ozturk, and
  Alkan]{bingol2021gatekeeper}
Z{\"u}lal Bing{\"o}l, Mohammed Alser, Onur Mutlu, Ozcan Ozturk, and Can Alkan.
\newblock {GateKeeper-GPU: Fast and accurate pre-alignment filtering in short
  read mapping}.
\newblock In \emph{2021 IEEE International Parallel and Distributed Processing
  Symposium Workshops (IPDPSW)}, pages 209--209. IEEE, 2021.

\bibitem[Fei et~al.(2018)Fei, Dan, Lina, Xin, and Chunlei]{Fei2018}
Xia Fei, Zou Dan, Lu~Lina, Man Xin, and Zhang Chunlei.
\newblock {FPGASW: Accelerating Large-Scale Smith--Waterman Sequence Alignment
  Application with Backtracking on FPGA Linear Systolic Array}.
\newblock \emph{Interdisciplinary Sciences: Computational Life Sciences},
  10\penalty0 (1):\penalty0 176--188, Mar 2018.
\newblock ISSN 1867-1462.
\newblock \doi{10.1007/s12539-017-0225-8}.
\newblock URL \url{https://doi.org/10.1007/s12539-017-0225-8}.

\bibitem[Haghi et~al.(2023)Haghi, Marco-Sola, Alvarez, Diamantopoulos,
  Hagleitner, and Moreto]{HAGHI202339}
Abbas Haghi, Santiago Marco-Sola, Lluc Alvarez, Dionysios Diamantopoulos,
  Christoph Hagleitner, and Miquel Moreto.
\newblock {WFA-FPGA: An efficient accelerator of the wavefront algorithm for
  short and long read genomics alignment}.
\newblock \emph{Future Generation Computer Systems}, 149:\penalty0 39--58,
  2023.
\newblock ISSN 0167-739X.
\newblock \doi{https://doi.org/10.1016/j.future.2023.07.008}.
\newblock URL
  \url{https://www.sciencedirect.com/science/article/pii/S0167739X2300256X}.

\bibitem[Ahmed et~al.(2019)Ahmed, L{\'e}vy, Ren, Mushtaq, Bertels, and
  Al-Ars]{Ahmed2019}
Nauman Ahmed, Jonathan L{\'e}vy, Shanshan Ren, Hamid Mushtaq, Koen Bertels, and
  Zaid Al-Ars.
\newblock {GASAL2: a GPU accelerated sequence alignment library for
  high-throughput NGS data}.
\newblock \emph{BMC Bioinformatics}, 20\penalty0 (1):\penalty0 520, Oct 2019.
\newblock ISSN 1471-2105.
\newblock \doi{10.1186/s12859-019-3086-9}.
\newblock URL \url{https://doi.org/10.1186/s12859-019-3086-9}.

\bibitem[Lindegger et~al.(2022)Lindegger, Cali, Alser, G{\'o}mez-Luna, and
  Mutlu]{lindegger2022algorithmic}
Jo{\"e}l Lindegger, Damla~Senol Cali, Mohammed Alser, Juan G{\'o}mez-Luna, and
  Onur Mutlu.
\newblock {Algorithmic improvement and GPU acceleration of the GenASM
  algorithm}.
\newblock In \emph{2022 IEEE International Parallel and Distributed Processing
  Symposium Workshops (IPDPSW)}, pages 162--162. IEEE, 2022.

\bibitem[Diab et~al.(2023)Diab, Nassereldine, Alser, G{\'o}mez~Luna, Mutlu, and
  El~Hajj]{diab2023framework}
Safaa Diab, Amir Nassereldine, Mohammed Alser, Juan G{\'o}mez~Luna, Onur Mutlu,
  and Izzat El~Hajj.
\newblock {A framework for high-throughput sequence alignment using real
  processing-in-memory systems}.
\newblock \emph{Bioinformatics}, 39\penalty0 (5):\penalty0 btad155, 2023.

\bibitem[Mao et~al.(2022{\natexlab{b}})Mao, Alser, Sadrosadati, Firtina,
  Baranwal, Cali, Manglik, Alserr, and Mutlu]{mao2022genpip}
Haiyu Mao, Mohammed Alser, Mohammad Sadrosadati, Can Firtina, Akanksha
  Baranwal, Damla~Senol Cali, Aditya Manglik, Nour~Almadhoun Alserr, and Onur
  Mutlu.
\newblock {Genpip: In-memory acceleration of genome analysis via tight
  integration of basecalling and read mapping}.
\newblock In \emph{2022 55th IEEE/ACM International Symposium on
  Microarchitecture (MICRO)}, pages 710--726. IEEE, 2022{\natexlab{b}}.

\bibitem[Gudur et~al.(2022)Gudur, Maheshwari, Acharyya, and Shafik]{9520260}
Venkateshwarlu~Yellaswamy Gudur, Sidharth Maheshwari, Amit Acharyya, and Rishad
  Shafik.
\newblock {An FPGA Based Energy-Efficient Read Mapper With Parallel Filtering
  and In-Situ Verification}.
\newblock \emph{IEEE/ACM Transactions on Computational Biology and
  Bioinformatics}, 19\penalty0 (5):\penalty0 2697--2711, 2022.
\newblock \doi{10.1109/TCBB.2021.3106311}.

\bibitem[Chen et~al.(2021)Chen, Chang, Yang, and Chiueh]{9320533}
Yen-Lung Chen, Bo-Yi Chang, Chia-Hsiang Yang, and Tzi-Dar Chiueh.
\newblock {A High-Throughput FPGA Accelerator for Short-Read Mapping of the
  Whole Human Genome}.
\newblock \emph{IEEE TPDS}, 32\penalty0 (6):\penalty0 1465--1478, 2021.
\newblock \doi{10.1109/TPDS.2021.3051011}.

\bibitem[Vasimuddin et~al.(2019)Vasimuddin, Misra, Li, and Aluru]{8820962}
Md. Vasimuddin, Sanchit Misra, Heng Li, and Srinivas Aluru.
\newblock {Efficient Architecture-Aware Acceleration of BWA-MEM for Multicore
  Systems}.
\newblock In \emph{2019 IEEE International Parallel and Distributed Processing
  Symposium (IPDPS)}, pages 314--324, 2019.
\newblock \doi{10.1109/IPDPS.2019.00041}.

\end{thebibliography}
